\documentclass[amsfonts,amssymb]{revtex4}

\usepackage{amsmath}
\usepackage{float}
\usepackage[dvips]{graphicx}
\usepackage{xcolor,ulem} 
\usepackage{amsfonts}
\usepackage{amsbsy}
\usepackage{amssymb}
\usepackage{verbatim}

\begin{document}
\preprint{}

\title{Statistical mechanics for complex systems: On the structure of $q$-triplets\footnote{Invited contribution to the Proceedings of the 31st International Colloquium on Group Theoretical Methods in Physics (Rio de Janeiro, 2016).}} 



\author{Constantino Tsallis}
\affiliation{Centro Brasileiro de Pesquisas Fisicas and National Institute of Science and Technology for Complex Systems, Rua Xavier Sigaud 150, 22290-180 Rio de Janeiro, Brazil}
\affiliation{Santa Fe Institute, 1399 Hyde Park Road, Santa Fe, New Mexico 87501, USA  \\tsallis@cbpf.br
}



\begin{abstract}
A plethora of natural, artificial and social complex systems exists which violate the basic hypothesis (e.g., ergodicity) of Boltzmann-Gibbs (BG) statistical mechanics.  Many of such cases can be satisfactorily handled by introducing nonadditive entropic functionals, such as  $S_q\equiv k\frac{1-\sum_{i=1}^W p_i^q}{q-1} \; \Bigl(q \in {\cal R}; \, \sum_{i=1}^W p_i=1 \Bigr)$, 
with $S_1=S_{BG}\equiv -k\sum_{i=1}^W p_i \ln p_i$. Each class of such systems can be characterized by a set of values  $\{q\}$, directly corresponding to its various physical/dynamical/geometrical properties. A most important subset is usually referred to as the $q$-triplet, namely $(q_{sensitivity}, q_{relaxation}, q_{stationary\,state})$, defined in the body of this paper. In the BG limit we have   $q_{sensitivity}=q_{relaxation}=q_{stationary\,state}=1$. For a given class of complex systems, the set $\{q\}$ contains only a few independent values of $q$, all the others being functions of those few. An illustration of this structure was given in 2005 [Tsallis, Gell-Mann and Sato, Proc. Natl. Acad. Sc. USA {\bf 102}, 15377; TGS]. This illustration enabled a satisfactory analysis of the Voyager 1 data on the solar wind. But the general form of these structures still is an open question. This is so, for instance, for the challenging $q$-triplet associated with the edge of chaos of the logistic map. We introduce here a transformation which sensibly generalizes the TGS one, and which might constitute an important step towards the general solution. 
\end{abstract}

\maketitle
\section{Introduction}
The pillars of contemporary theoretical physics may be considered to be Newtonian, quantum and relativistic mechanics, Maxwell electromagnetism, and Boltzmann-Gibbs (BG) statistical mechanics (microscopic theory consistent with thermodynamics). Statistical mechanics is in turn grounded upon {\it electromechanics} (meaning by this the set of all mechanics and electromagnetism) and theory of probabilities. The BG theory can be formally constructed by adopting the BG entropic functional $S_{BG}=-k\sum_{i=1}^W p_i \ln p_i$, with $\sum_{i=1}^W p_i=1$, $k$ being a conventional positive constant (usually taken to be the Boltzmann constant $k_B$). This hypothesis is known to be fully satisfactory for dynamical systems satisfying simple properties such as ergodicity. For more complex systems, the BG entropy can be inadequate, even plainly misleading. When this happens, must we abandon the statistical mechanical approach? It was advanced in 1988 \cite{Tsallis1988} that this is not necessary. Indeed, it suffices to consider entropic functionals different from $S_{BG}$, and reconstruct statistical mechanics on more general grounds. The so called {\it nonextensive statistical mechanics} follows along this path, based on the entropy $S_q = k\frac{1-\sum_{i=1}^W p_i^q}{q-1} \; (q \in {\cal R};\,S_1=S_{BG})$.  It can be easily verified that, if $A$ and $B$ are any two probabilistically independent systems (i.e., $p_{ij}^{A+B}=p_i^A p_j^B$), $\frac{S_q(A+B)}{k}=\frac{S_q(A)}{k}+\frac{S_q(B)}{k}+(1-q)\frac{S_q(A)}{k}\frac{S_q(B)}{k}$. In other words, $S_q$ is {\it nonadditive} for $q \ne 1$, in contrast with $S_{BG}$ which is {\it additive}.

The optimization of $S_q$ under appropriate constraints yields distributions such as the $q$-exponential one $p_q(x) \propto [1-(1-q) \beta x]^{1/(1-q)} \equiv e_q^{-\beta x}$ or the $q$-Gaussian one $p_q(x) \propto e_q^{-\beta x^2}$ (see \cite{Tsallis2009} for an introductory text). This and similar generalizations of the BG statistical mechanics have been shown to provide uncountable predictions, verifications and applications in natural, artificial and social complex systems. A regularly updated bibliography as well as selected theoretical, experimental, observational, and computational papers can be seen at  http://tsallis.cat.cbpf.br/biblio.htm  Among recent applications we may mention the experimental validation~\cite{CombeRichefeuStasiakAtman2015} (accomplished in granular matter) of a 20-year-old prediction, the emergence of neat $q$-statistical behavior in high-energy collisions at LHC/CERN along 14 experimental decades (see \cite{WongWilk2013} for instance), a notable numerical discovery in the celebrated standard map \cite{TirnakliBorges2016}, and the connection with networks (see \cite{BritoSilvaTsallis2016} for instance).

\section{$q$-triplets}

The solution of the differential equation
\begin{equation}
\frac{dy}{dx}= a_1 y \;\;(y(0)=1)
\end{equation}
is given by $y=e^{a_1 x}$.
The solution of the more general equation
\begin{equation}
\frac{dy}{dx}= a_q y^q \;\;(y(0)=1)
\label{diffeq}
\end{equation}
is given by $y=e_q^{a_qx}$.
These facts in the realm of nonextensive statistical mechanics suggested a conjecture in 2004 \cite{Tsallis2004}, namely that there could exist in nature $q$-triplets as indicated in Table \ref{table} and \cite{Tsallis2016}. The first verification of the conjecture was done in 2005 by the NASA researchers Burlaga and Vinas in the solar wind \cite{BurlagaVinas2005}.  
\begin{table}[htbp]
\begin{center}
\begin{tabular}{c||c|c|c||}
                                                            &  $x$         & $a$                                       &$y(x)$                                                                             \\
[1mm] \hline\hline
& & & \\
Stationary state distribution                &  $E_i$     & $-\beta$                                 & $Z_{q_{stationary\,state}} \, p(E_i)=e_{q_{stationary\,state}}^{-\beta_{q_{stationary\,state}}\, E_i}$        \\
[3mm] \hline
& & & \\
Sensitivity to the initial conditions       &  $t$         & $\lambda_{q_{sensitivity}}$              & \;\;\;\;\;\;\;\;\;\;\;\;\;\;\;$\xi(t)=e_{q_{sensitivity}}^{ \, \lambda_{q_{sensitivity}}\,t}$             \\
[3mm] \hline 
& & & \\
Typical relaxation of observable $O$   &$t$           &$-1/\tau_{q_{relaxation}}$                  & \;\;\;\;\;\;\;\;\;\;\;\;\;\;\;$\Omega(t) \equiv \frac{O(t)-O(\infty)}{O(0)-O(\infty)}=e_{q_{relaxation}}^{-t/\tau_{q_{relaxation}}}$                   \\
[3mm] \hline \hline
\end{tabular}
\end{center}
\caption{Three possible physical interpretations of Eq. (\ref{diffeq}) within nonextensive statistical mechanics. In the BG limit we have $q_{sensitivity}=q_{relaxation}=q_{stationary\,state}=1$. For one dimensional dynamical systems it is $q_{entropy\,production}=q_{sensitivity}$, where $q_{entropy\,production}$ denotes the index $q$ for which $S_q$ increases linearly with time $t$. From \cite{Tsallis2016}.}
\label{table}
\end{table}
Since then a plethora of $q$-triplets and directly related quantities have been found in solar plasma \cite{BurlagaNess2013,PavlosKarakatsanisXenakis2012,KarakatsanisPavlosXenakis2013,PavlosIliopoulosZastenkerZelenyiKarakatsanisRiazantsevaXenakisPavlos2015}, the ozone layer \cite{FerriSavioPlastino2010}, El Ni\~no/Southern Oscillations \cite{FerriFigliolaRosso2012}, geological faults \cite {FreitasFrancaScherrerVilarSilva2013}, finance\cite{PavlosKarakatsanisXenakisPavlosIliopoulosSarafopoulos2014,IliopoulosPavlosMagafasKarakatsanisXenakisPavlos2015}, DNA sequence \cite{PavlosKarakatsanisIliopoulosPavlosXenakisClarkDukeMonos2015}, logistic map (see \cite{TsallisPlastinoZheng1997,LyraTsallis1998,Lyra1998,BaldovinRobledo2004a,BaldovinRobledo2004b,MayoralRobledo2004a,MayoralRobledo2004b,TirnakliBeckTsallis2007,TirnakliTsallisBeck2009,Grassberger2009,AnanosBaldovinTsallis2005,LuqueLacasaRobledo2012}), and elsewhere \cite{Tsallis2006,SuyariWada2007}.

\section{Connections between $q$-indices}

Some very basic points can be addressed at this stage: How many indices $q$ can be systematically defined? How many of them are independent?  Through what relations can all the others be calculated? To what specific physical/mathematical/probabilistic/dynamical property is each of them associated? 

As we shall see, there are many more than three relevant $q$-indices. Nevertheless, the $q$-triplet plays a kind of guiding role in questions such as what is the correct entropy to be used, at what rhythm it relaxes to a stationary state, and how this stationary state can be characterized. Consistently, in the BG limit all the indices $q$ are expected to be equal among them and equal to unity.

Inspired by the specific values for the $q$-triplet observed by NASA \cite{BurlagaVinas2005}, a path was developed in \cite{TsallisGellMannSato2005}. Two self-dual transformations admitting $q=1$ as a fixed point were introduced, namely the {\it additive duality} $q \to 2-q$ and the {\it multiplicative duality} $q \to 1/q$. These simple transformations had already appeared in various contexts in nonextensive statistical mechanics (see \cite{Tsallis2009} and references therein). The novelty in \cite{TsallisGellMannSato2005} is that they were used to systematically construct a mathematical structure, which we describe in what follows. We first define the transformations $\mu$ and $\nu$: 
\begin{equation}
\mu \;\to\; q_2(q)=2-q \;\to\; \frac{1}{1-q_2(q)}=\frac{1}{q-1}\,,
\end{equation}
\begin{equation}
\nu \;\to\; q_0(q)=\frac{1}{q} \;\to\; \frac{1}{1-q_0(q)}=\frac{1}{q-1} +1 \,.
\end{equation}
The subindices 2 and 0 will become clear soon. We straightforwardly verify $\mu^2=\nu^2=1$, $\nu\mu=(\mu\nu)^{-1}$. Also, we can analogously define $(\mu\nu)^m$ and $(\nu\mu)^n$ with integer numbers$(m,n)$. This set of transformations enables (see \cite{TsallisGellMannSato2005,Tsallis2009})  the definition of a simple structure (hereafter referred to as the TGS structure). The NASA $q$-triplet for the solar wind found an elegant description within this structure, as shown later on in this paper. Not so the logistic-map edge-of-chaos $q$-triplet, and others. As a possible way out of this limitation, a generalization of the TGS  structure was proposed in \cite{Tsallis2016}, which we review now.

Let us consider the following transformation:
\begin{equation}
q_a(q)=\frac{(a+2) -aq}{a-(a-2)q}  \;\;(a \in {\cal R})\,,
\label{qdualitynew}
\end{equation}
or, equivalently,
\begin{equation}
\frac{1}{1-q_a(q)}=\frac{1}{q-1}  + 1-\frac{a}{2}\,,
\label{qduality2}
\end{equation}
or, even,
\begin{equation}
\frac{2}{2-a}\frac{1}{1-q_a(q)}=\frac{2}{2-a}\frac{1}{q-1}  + 1\,.
\label{qduality3}
\end{equation}

We straightforwardly verify that $q_2=2-q$ ({\it additive duality}) and $q_0=1/q$ ({\it multiplicative duality}) \cite{TsallisGellMannSato2005,Tsallis2009,Tsallis2009b,Tsallis2012}. Also, we generically verify {\it selfduality}, i.e., $q_a(q_a(q))=q \,, \forall (a,q)$, as well as the BG fixed point, i.e., $q_a(1)=1 \,, \forall a$: See the figure in \cite{Tsallis2016}.
The duality \eqref{qdualitynew} is in fact a quite general ratio of linear functions of $q$  which satisfies these two important properties (selfduality and BG fixed point). It transforms biunivocally the interval $[1,-\infty)$ into the interval $[1,\frac{a}{a-2}]$.
Moreover, for $a=3$ and $a=5$ we recover respectively $q_3=\frac{5-3q}{3-q}$ \cite{NelsonUmarov2010} and $q_5=\frac{7-5q}{5-3q}$ \cite{HanelThurnerTsallis2009}. 

Let us combine now two\footnote{Of course, it is also possible to combine, along similar lines, three or more such transformations.} transformations of the type  \eqref{qdualitynew} (or, equivalently, \eqref{qduality2}):
\begin{equation}
\mu \; \to \; q_a(q)=\frac{(a+2) -aq}{a-(a-2)q} \; \to \; \frac{1}{1-q_a(q)}=\frac{1}{q-1}  + 1-\frac{a}{2} \,,
\label{transform1}
\end{equation}
and
\begin{equation}
\nu \; \to \; q_b(q)=\frac{(b+2) -bq}{b-(b-2)q} \; \to \; \frac{1}{1-q_b(q)}=\frac{1}{q-1}  + 1-\frac{b}{2} \,,
\label{transform2}
\end{equation}
with $b \ne a$. It follows that
\begin{equation}
\mu \nu  \; \to \; q_a(q_b(q))=\frac{(b-a) -(b-a-2)q}{(b-a+2)-(b-a)q} \; \to \; \frac{1}{1-q_a(q_b(q))}=\frac{1}{1-q}  + \frac{b-a}{2} \,,
\label{transform3}
\end{equation}
and
\begin{equation}
\nu \mu  \; \to \; q_b(q_a(q))=  \frac{(a-b) -(a-b-2)q}{(a-b+2)-(a-b)q} \; \to \; \frac{1}{1-q_b(q_a(q))}=\frac{1}{1-q}  + \frac{a-b}{2} \,,
\label{transform4}
\end{equation}
with $\mu^2=\nu^2=1$, $\nu \mu=(\mu \nu)^{-1}$, and $q_a(q_a(q))=q \,, \forall (a,q)$. 

For integer values of $m$ and $n$, we can straightforwardly establish
\begin{eqnarray}
(\mu \nu)^m  \; &\to& \; q_{a,b}^{(m)}(q) \equiv q_a(q_b(q_a(q_b(...))))=\frac{m(b-a) -[m(b-a)-2]q}{[m(b-a)+2]-m(b-a)q} \\ \; &\to& \; \frac{1}{1-q_{a,b}^{(m)}(q)} =\frac{1}{1-q_a(q_b(q_a(q_b(...))))}=\frac{1}{1-q}  + m\frac{b-a}{2} \,,
\label{algebra1}
\end{eqnarray}
and
\begin{eqnarray}
(\nu \mu)^n  \; &\to& \; q_{b,a}^{(n)}(q) \equiv q_b(q_a(q_b(q_a(...))))=  \frac{n(a-b) -[n(a-b)-2]q}{[n(a-b)+2]-n(a-b)q} \\ \; &\to& \; \frac{1}{1-q_{b,a}^{(n)}(q)} = \frac{1}{1-q_b(q_a(q_b(q_a(...))))}=\frac{1}{1-q}  +n \frac{a-b}{2} \,.
\label{algebra2}
\end{eqnarray}
As we see, $q_{a,b}^{(1)}=q_a(q_b(q))$ and $q_{b,a}^{(1)}=q_b(q_a(q))$.

For $a \ne b$ and any integer values for $(m,n)$, the above general relations can be conveniently rewritten as follows:
\begin{eqnarray}
\frac{2}{b-a} \frac{1}{1-q_{a,b}^{(m)}(q)} =\frac{2}{b-a}\frac{1}{1-q}  + m \;\;\;(m=0,\pm 1,\pm2, ...) \,,
\end{eqnarray}
and
\begin{eqnarray}
\frac{2}{a-b} \frac{1}{1-q_{b,a}^{(n)}(q)} =\frac{2}{a-b}\frac{1}{1-q}  + n \;\;\;(n=0,\pm 1,\pm2, ...) \,.
\end{eqnarray}
For $m=n=1$ and $(a,b)=(2,0)$ we recover the simple transformations $q_{2,0}^{(1)}=2-\frac{1}{q}$ (see Eq. (7) in \cite{MoyanoTsallisGellMann2006}, and footnote in page 15378 of \cite{TsallisGellMannSato2005}) and $q_{0,2}^{(1)}=\frac{1}{2-q}$.

We can also check that, with $m=0,\pm 1,\pm 2, ...$, $(\mu \nu)^m \mu$ and $\nu(\mu \nu)^m$ correspond respectively to 
\begin{eqnarray}
\frac{2}{b-a} \frac{1}{1-q_{a,b}^{(m, \mu)}(q)} -\frac{2-a}{2(b-a)}=- \Bigl[\frac{2}{b-a} \frac{1}{1-q} -\frac{2-a}{2(b-a)} \Bigr] - m \,,
\label{algebra3}
\end{eqnarray}
and
\begin{eqnarray}
\frac{2}{b-a} \frac{1}{1-q_{a,b}^{(\nu, m)}(q)} -\frac{2-b}{2(b-a)}=  -\Bigl[\frac{2}{b-a} \frac{1}{1-q} -\frac{2-b}{2(b-a)}\Bigr] + m \,.
\label{algebra4}
\end{eqnarray}

Analogously we can check that, with $n=0,\pm 1,\pm 2, ...$, $(\nu \mu)^n\nu$ and  $\mu(\nu \mu)^n$ correspond respectively to
\begin{eqnarray}
\frac{2}{a-b} \frac{1}{1-q_{b,a}^{(n, \nu)}(q)} -\frac{2-b}{2(a-b)}=- \Bigl[\frac{2}{a-b} \frac{1}{1-q} -\frac{2-b}{2(a-b)} \Bigr] - n \,,
\label{algebra5}
\end{eqnarray}
and
\begin{eqnarray}
\frac{2}{a-b} \frac{1}{1-q_{b,a}^{(\mu, n)}(q)} -\frac{2-a}{2(a-b)}=  -\Bigl[\frac{2}{b-a} \frac{1}{1-q} -\frac{2-a}{2(a-b)}\Bigr] + n \,.
\label{algebra6}
\end{eqnarray}

As we see, the structures that are involved exhibit some degree of complexity. Let us therefore summarize the frame within which we are working. If we have an unique parameter (noted $a$) to play with, we can only transform $q$ through Eq. (\ref{qdualitynew}). If we have two parameters (noted $a$ and $b$) to play with, we can transform $q$ in several ways, namely through Eqs. (\ref{algebra1}), (\ref{algebra2}), (\ref{algebra3}), (\ref{algebra4}), (\ref{algebra5}) and (\ref{algebra6}), with $m=0,\pm 1,\pm 2, ...$ and $n=0,\pm 1,\pm 2, ...$; the cases $m=0$ and $n=0$ recover respectively Eqs. (\ref{transform1}) and (\ref{transform2}). The particular choice $(a,b)=(2,0)$ recovers the TGS structure introduced in \cite{TsallisGellMannSato2005,Tsallis2009,Tsallis2009b,Tsallis2012}. Also, the particular choice $(a,b)=(-1,0)$ within the transformation (\ref{transform3}) recovers the transformation $q \to \frac{1+q}{3-q}$, which plays a crucial role in the $q$-generalized Central Limit Theorem \cite{UmarovTsallisSteinberg2008}; coincidentally (or not), the relation $b-a=1$ recovers the $\gamma=1/2$ case of Eq. (32) of \cite{Tsallis2016} (see also \cite{RuizTsallis2012,Touchette2013,RuizTsallis2013}).

To make the approach introduced in \cite{Tsallis2016} even more powerful, we may introduce now the 
{\it most general self-dual ratio of linear functions of $q$, which has the $q=1$ fixed point}. It is given by
\begin{equation}
q_{a_1,a_2}(q) = \frac{a_1-a_2 q}{a_2-(2a_2-a_1)q} \;\;(a_1 \in {\cal R}; \, a_2 \in {\cal R})\,,
\label{generalizedtransform}
\end{equation}
or, equivalently,
\begin{equation}
\frac{1}{1-q_{a_1,a_2}(q)} = \frac{1}{q-1}+1+ \frac{a_2}{a_2-a_1} \,,
\end{equation}
or, even,
\begin{equation}
\frac{a_2-a_1}{2a_2-a_1}\frac{1}{1-q_{a_1,a_2}(q)} =\frac{a_2-a_1}{2a_2-a_1} \frac{1}{q-1}+1 \,.
\label{generaltransform}
\end{equation}

The particular case
\begin{equation}
(a_1,a_2) =(a+2,a) 
\label{particular}
\end{equation}
recovers the transformation introduced in Eq. (\ref{qdualitynew}) \cite{Tsallis2016}. All the steps from Eq. (\ref{transform1}) to Eq. (\ref{algebra6}) can easily be generalized, involving now four parameters, $(a_1,a_2,b_1,b_2)$, instead of only two, $(a,b)$. It becomes clear that the 4-parameter structure that can be constructed with the transformation (\ref{generaltransform}) remains isomorphic to the set {\cal Z} of integer numbers. Of course, to go from the 4-parameter structure to the 2-parameter structure we need to assume also, analogously to Eq. (\ref{particular}), that $(b_1,b_2) =(b+2,b)$.
 
\section{Some final remarks}
Essentially, we reproduce here the final remarks in \cite{Tsallis2016}.
The data observed in \cite{BurlagaVinas2005} for the solar wind are consistent with the $q$-triplet \cite{TsallisGellMannSato2005} $(q_{sensitivity}, q_{stationary\,state},q_{relaxation})=(-0.5,7/4,4)$.

If we identify, in Eq. (\ref{transform3}), $(q,q_{a,b}^{(1)}) \equiv (q_{sensitivity},q_{relaxation})$ we can verify that, for $a-b=2$, the data are consistently recovered. Moreover, if we use once again Eq. (\ref{transform3}) and $a-b=2$, but identifying now $(q,q_{a,b}^{(1)}) \equiv (q_{relaxation},q_{stationary\,state})$, once again the data are consistently recovered. The particular case $(a,b)=(2,0)$ was first proposed in \cite{TsallisGellMannSato2005}. In other words, it is possible to consider this $q$-triplet as having only one independent value, say $q_{sensitivity}$; from this value we can calculate $q_{relaxation}$ by using  Eq. (\ref{transform3}); and from $q_{relaxation}$ we can calculate $q_{stationary\,state}$ by using once again Eq. (\ref{transform3}). This discussion can be summarized as follows:
\begin{equation}
\frac{1}{1-q_{sensitivity}}-\frac{1}{1-q_{relaxation}}  = \frac{1}{1-q_{relaxation}}-\frac{1}{1-q_{stationary\,state}} =   \frac{a-b}{2}=1 \,.
\end{equation}
It is occasionally convenient to use the $\epsilon$-triplet defined as $(\epsilon_{sensitivity},\epsilon_{stationary\,state},\epsilon_{relaxation})= (1-q_{sensitivity},1-q_{stationary\,state},1-q_{relaxation})$. Let us mention that an amazing set of relations was found among these by \cite{Baella2008}, namely 
\begin{eqnarray}
\epsilon_{stationary\,state} = \frac{\epsilon_{sensitivity} + \epsilon_{relaxation}}{2} \,,    \\
\epsilon_{sensitivity}=\sqrt{\epsilon_{stationary\,state} \, \epsilon_{relaxation}}  \,, \\
\epsilon_{relaxation}^{-1} =\frac{\epsilon_{sensitivity}^{-1} + \epsilon_{stationary\,state}^{-1}}{2}   \,.
\end{eqnarray}
The emergence of the three Pythagorean means in this specific $q$-triplet remains still today enigmatic. One could advance that these relations hide some unexpected symmetry, but its nature remains today completely unrevealed.

Let us now focus on a different system, namely the well known logistic map at its edge of chaos (also referred to as the Feigenbaum point). The numerical data for this map yield the $q$-triplet  $(q_{sensitivity}, q_{stationary\,state},q_{relaxation})=(0.244487701...,1.65 \pm 0.05,2.249784109...)$ \cite{LyraTsallis1998,MouraTirnakliLyra2000,Grassberger2005,Robledo2006,TirnakliTsallisBeck2009}.

An heuristic relation has been found \cite{Baella2010} between these three values, namely 
(using $\epsilon \equiv 1-q$)
\begin{equation}
\epsilon_{sensitivity} + \epsilon_{relaxation} = \epsilon_{sensitivity} \, \epsilon_{stationary\,state} \,.
\label{baella2}
\end{equation}
Indeed, this relation straightforwardly implies
\begin{equation}
q_{stationary\,state}=\frac{q_{relaxation}-1}{1-q_{sensitivity}} \,.
\end{equation}
Through this relation we obtain $q_{stationary\,state}=1.65424...$ which is perfectly compatible with $1.65 \pm 0.05$. 
In the generalized structure that we have developed here above we have five free parameters $(q,a_1,a_2,b_1,b_2)$ (or only three free parameters $(q,a,b)$ in the more restricted version presented in \cite{Tsallis2016}) in addition to the integer numbers $(m,n)$. It is therefore trivial to make analytical identifications with $(q_{sensitivity}, q_{stationary\,state},q_{relaxation})$ such that Eq. (\ref{baella2}) is satisfied. 

The real challenge, however, is to find a general theoretical frame within which such identifications (and, through the freedom associated with $(m,n)$, infinitely many more, related to physical quantities) become established on a clear basis, and not only through conjectural possibilities; as a simple illustration of such $q$ indices being associated to specific properties, we may mention the relation \cite{CelikogluTirnakliQueiros2010,BakarTirnakli2010,CelikogluTirnakli2012} $q_{stationary\,state}=\frac{\tau+2}{\tau}$, hence $q_{stationary\,state}-1=2(q_{avalanche\,size}-1)$ with $\tau \equiv 1/(q_{avalanche\,size}-1)$.
Such a frame of systematic identifications remains up to now elusive and certainly constitutes a most interesting open question. Along this line, a connection that might reveal promising is that, if we assume that $q$ is a complex number (see, for instance, \cite{WilkWlodarczyk2015,AzmiCleymans2015}), then Eq. (\ref{qdualitynew}) corresponds to nonsingular [with $(a+2)(a-2)-a^2=-4 \ne 0\,,\forall a$] Moebius transformations, which form the Moebius group, defining an automorphism of the Riemann sphere.

\section*{Acknowledgments}
I am deeply indebted to Piergiulio Tempesta. Indeed, during a long and fruitful conversation with him about the present context focusing on the structure and use of $q$-triplets based on transformation (\ref{qdualitynew}), he thought of generalizing it into transformation (\ref{generalizedtransform}).  Also, partial financial support by CNPq and Faperj (Brazilian agencies) and by the John Templeton Foundation (USA) is gratefully acknowledged.

\end{document}